\documentclass[twoside,twocolumn,9pt]{article}
\usepackage{extsizes}
\usepackage[super,sort&compress,comma]{natbib} 
\usepackage[version=3]{mhchem}
\usepackage[left=1.5cm, right=1.5cm, top=1.785cm, bottom=2.0cm]{geometry}
\usepackage{balance}
\usepackage{times,mathptmx}
\usepackage{sectsty}
\usepackage{graphicx} 
\usepackage{lastpage}
\usepackage[format=plain,justification=normal,singlelinecheck=false,font={stretch=1.125,small,sf},labelfont=bf,labelsep=space]{caption}
\usepackage{float}
\usepackage{fnpos}
\usepackage[english]{babel}
\usepackage{array}
\usepackage{droidsans}
\usepackage{charter}
\usepackage[T1]{fontenc}
\usepackage[usenames,dvipsnames]{xcolor}
\usepackage{setspace}
\usepackage[compact]{titlesec}
\usepackage{hyperref}

\definecolor{cream}{RGB}{222,217,201}

\begin{document}

\thispagestyle{plain}


\makeFNbottom
\makeatletter
\renewcommand\LARGE{\@setfontsize\LARGE{15pt}{17}}
\renewcommand\Large{\@setfontsize\Large{12pt}{14}}
\renewcommand\large{\@setfontsize\large{10pt}{12}}
\renewcommand\footnotesize{\@setfontsize\footnotesize{7pt}{10}}
\makeatother

\renewcommand{\thefootnote}{\fnsymbol{footnote}}
\renewcommand\footnoterule{\vspace*{1pt}%
\hrule width 3.6in height 0.4pt \color{black}\vspace*{5pt}} 
\setcounter{secnumdepth}{5}

\makeatletter 
\renewcommand\@biblabel[1]{#1}            
\renewcommand\@makefntext[1]%
{\noindent\makebox[0pt][r]{\@thefnmark\,}#1}
\makeatother 
\renewcommand{\figurename}{\small{Fig.}~}
\sectionfont{\sffamily\Large}
\subsectionfont{\normalsize}
\subsubsectionfont{\bf}
\setstretch{1.125} 
\setlength{\skip\footins}{0.8cm}
\setlength{\footnotesep}{0.25cm}
\setlength{\jot}{10pt}
\titlespacing*{\section}{0pt}{4pt}{4pt}
\titlespacing*{\subsection}{0pt}{15pt}{1pt}


\makeatletter 
\newlength{\figrulesep} 
\setlength{\figrulesep}{0.5\textfloatsep} 

\makeatother

\twocolumn[
  \begin{@twocolumnfalse}
\sffamily
\noindent\huge{\textbf{A versatile method to generate multiple types of micropatterns}} \\
\vspace{0.05cm} \\
\noindent\large{F. J. Segerer,\textit{$^{a}$} P. J. F. R\"ottgermann,\textit{$^{a}$} S. Schuster,\textit{$^{b}$} A. Piera Alberola,\textit{$^{a}$} S. Zahler,\textit{$^{b}$} and J. O. R\"adler\textit{$^{\ast a}$}} \\
\vspace{0.3cm} \\
\noindent\normalsize{Micropatterning techniques have become an important tool for the study of cell behavior in controlled microenvironments. As a consequence, several approaches for the creation of micropatterns have been developed in recent years. However, the diversity of substrates, coatings and complex patterns used in cell science is so great that no single existing technique is capable of fabricating designs suitable for all experimental conditions. Hence, there is a need for patterning protocols that are flexible with regard to the materials used and compatible with different patterning strategies to create more elaborate setups. In this work, we present a versatile approach to micropatterning. The protocol is based on plasma treatment, protein coating, and a PLL-PEG backfill step, and produces homogeneous patterns on a variety of substrates. Protein density within the patterns can be controlled, and density gradients of surface-bound protein can be formed. Moreover, by combining the method with microcontact printing, it is possible to generate patterns composed of three different components within one iteration of the protocol. The technique is simple to implement and should enable cell science labs to create a broad range of complex and highly specialized microenvironments.} \\


\end{@twocolumnfalse} 
\vspace{0.6cm}

]


\rmfamily
\section*{}
\vspace{-1cm}


\footnotetext{\textit{$^{a}$~Faculty of Physics and Center for NanoScience, Ludwig-Maximilians-Universit\"at M\"unchen, Geschwister-Scholl-Platz 1, D-80539 Munich, Germany. Fax: +49 89 2180 3182; Tel: +49 89 2180 2438; E-mail: raedler@lmu.de}}
\footnotetext{\textit{$^{b}$~Department of Pharmacy - Center for Drug Research, Pharmaceutical Biology, Ludwig-Maximilians-Universit\"at M\"unchen, Butenandtstr. 5-13, D-81377 Munich, Germany }}





\section*{Introduction}
Micropatterning techniques have become an established tool for researchers interested in single-cell functions and dynamics \cite{Mahmud2009,Thery2005,Thery2006,Parker2002,Jiang2005,Yoon2012, Ferizi2015} and the collective behavior of small cell assemblies and tissues \cite{Rolli2012,Marel2014,Segerer2015}. Their significance for today's cell science arises from the fact that they provide direct control over the shape and functionality of the cell's environment on a microscopic scale.

How a cell adapts to the structure and composition of its microenvironment can give considerable insight into its intrinsic mechanical and functional properties \cite{Bischofs2008,Tee2015,Thery2006-2}. In addition, micropatterns can be exploited to actively manipulate cell behavior. For instance, it has been found that the size and geometry of the accessible area can alter and direct the axis of cell polarization and division \cite{Thery2005,Thery2006,Jiang2005}, the positions and orientation of pseudopodia \cite{Parker2002}, or the locations of junctions between adjacent cells \cite{Tseng2012}. Cell adhesion and migration depends in large part on the protein composition of a cell's surroundings \cite{Junker1987,Tomaselli1987,Eichinger2012}, but the relative density of adhesion sites \cite{Maheshwari1999,Maheshwari2000,Rajagopalan2004}, as well as density gradients of surface-bound proteins, can influence and bias its spreading and motion \cite{Smith2004,Liu2007}.

Micropatterning techniques should therefore provide for precise control over the shape of the cell's microenvironment, i.e. the distribution and concentration of surface-bound proteins. In addition, it should be compatible with a large variety of proteins and be capable of producing patterns composed of multiple protein species.

Most current micropatterning protocols are based on one of the following three approaches. Soft lithography in form of microcontact printing ($\mathrm{\mu CP}$) involves protein transfer from a polymeric stamp, while the remaining surface is often passivated by PEGylation \cite{Thery2009,Ruiz2007,Wilbur1996}. In photolithography, the properties of a surface or a precoated matrix are locally altered by photocleavage with a laser device or by exposing it to light through a photomask \cite{Azioune2009,Kim2010,Belisle2009,Nakanishi2004}. In plasma lithography, a surface that is partially protected by a shadow mask or stamp is modified/activated by exposure to a plasma (e.g. oxygen) \cite{Cheng2010,Junkin2011,Langowski2005,Tourovskaia2003,Picone2014,Kim2011}. Protein deposition in the latter protocols is mainly achieved by surface adsorption from an aqueous solution. Each of these approaches has its own specific advantages. $\mathrm{\mu CP}$ provides flexibility with respect to the molecules that are transferred to the surface, and does not require advanced or expensive equipment. Photolithography-based protocols produce very homogeneous patterns and have also been extended to enable formation of gradients in the surface-bound protein density \cite{Belisle2009}. Finally, for plasma-based approaches profit from the strong activation of the surface by the plasma exposure which can be exploited (i) directly to cause increased cell attachment on elsewise cell repellent substrates \cite{Junkin2011,Kim2011}, (ii) as a basis to spatially control polymer or protein deposition or conformation \cite{Langowski2005,Cheng2010} and (iii) to selectively remove a layer of protein or polymer \cite{Picone2014}. The plasma treatment itself is a robust and effective procedure, that provides fast working protocols and is applicable to a large variety of substrates. However, the diversity of substrates, coatings and patterns used in the field of cell science is so great that no single existing technique is capable of fabricating designs suitable for all experimental conditions. In particular, gradients in protein density or the accurate deposition of different proteins within a multi component pattern are often difficult to accomplish. Additionally, since micropatterning should ideally be accessible to a broad range of labs, patterning methods should preferably also be easy to handle and cost efficient. Therefore, simple working protocols that are adaptable to different experimental conditions such as proteins and substrates and can be combined with other patterning approaches to create more complex microenvironments can be expected to stimulate further progress in this field.

In this paper, we present an alternative plasma based and simple means of creating micropatterns on a broad range of substrates. The technique is based on plasma-induced patterning in combination with PEGylation and protein coating, and is therefore referred to here as microscale plasma-initiated protein patterning ($\mu$PIPP). It provides control over the final concentration of protein on the surface and produces homogeneous and stable patterns on various substrates such as glass, tissue culture polystyrene (tc-PS), cyclic olefine copolymers (COCs), and parylene C. We show that gradients on the surface-bound protein density can be generated via protein incubation within a chemotaxis chamber. Finally we combine the $\mu$PIPP protocol with $\mathrm{\mu CP}$ to create complex patterns consisting of up to three different components while providing accurate and adjacent relative positioning. The method presented in this paper should prove useful as a facile and versatile approach to the fabrication of a wide variety of micropatterns.

\begin{figure*}[ht!]
\centering
  \includegraphics[width=1\linewidth]{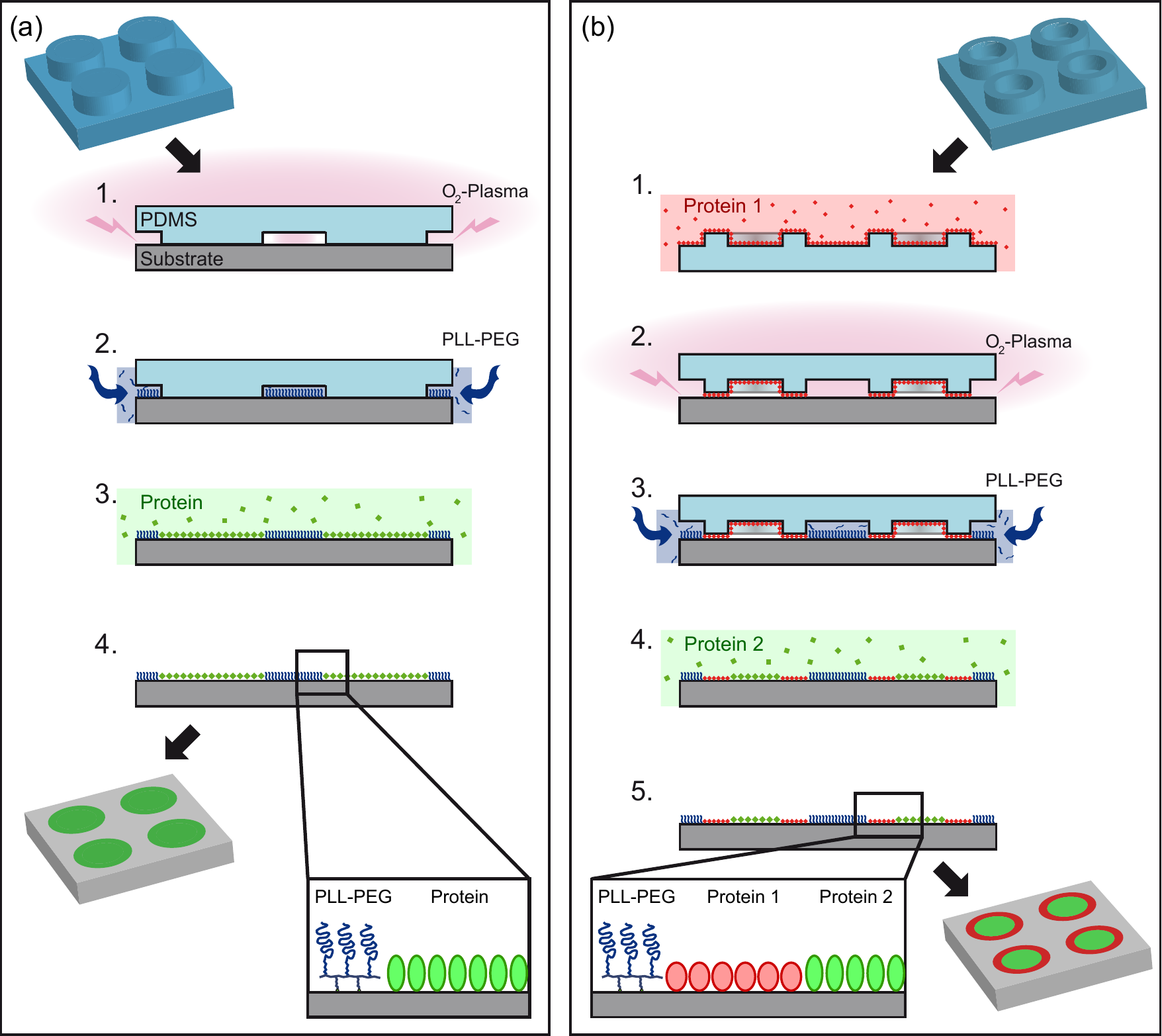}
  \caption{Patterning protocols. (a) Patterning procedure for conventional $\mu$PIPP: 1. The surface is partially covered by a PDMS stamp of the desired pattern and exposed to $\mathrm{O_2}$ plasma 2. A PLL-PEG solution is applied to the margins of the stamp and is drawn over the exposed surface by capillary action 3. The stamp is removed and the surface is incubated with the desired protein. (b) $\mu$PIPP combined with $\mu$CP: 1. Following UV-ozone activation (see Methods), the PDMS stamp is incubated with Protein 1. 2. The protein-coated stamp is inverted and Protein 1 is printed on the surface, which is simultaneously exposed to $\mathrm{O_2}$ plasma 3. PLL-PEG solution is applied to the stamp edge and drawn between stamp and surface by capillary action 4. The stamp is removed and the surface is incubated with Protein 2.}
  \label{fig:1}
\end{figure*}

\section*{Experimental}
\subsection*{Preparation of masters}
Masters for stamp preparation can be created by following established protocols (such as those provided by photoresist producers like MicroChem) or the protocol provided in section S1 of the Supplementary Material. Note that labs that do not have the means to create stamp masters can order them online (from HTS Resources, for example). Once prepared, each master can be used to make multiple stamps.

\subsection*{PDMS stamp preparation}
PDMS was prepared by mixing 10 parts silicone elastomer with 1 part cross linker (Sylgard Elastomer Kit, Dow Corning), and poured as a 1-3\;mm thick layer onto the master and degassed in a desiccator. The coated master was then cured overnight at a temperature of $\mathrm{50^{\circ}C}$.
   
\subsection*{Patterning}
The following sections describe the patterning protocols. The proteins used for patterning in this work were: fibrinogen labeled with Alexa-488, Alexa-594 or Alexa-647 (Life Technologies) respectively, fibronectin (YO Proteins) and laminin-1 (bio-techne). Note, however, that the technique may be used with a broad range of different proteins.

\subsubsection*{Conventional $\mu$PIPP:}
A PDMS stamp of the desired pattern was placed on the surface to be patterned (see Fig.~\ref{fig:1}a). The assembly was then exposed to $\mathrm{O_2}$ plasma at a pressure of 2 mbar in a plasma cleaner (Diener Femto) at 40\;W for 3\;min, thus activating the exposed parts of the surface. A droplet of a 2\;mg/ml PLL(20kDa)-g[3.5]-PEG(2kDa) (PLL-PEG) (SuSoS) solution in 10\;mM HEPES (pH 7.4) and 150\;mM NaCl was then placed at the edge of the stamp, and was drawn into the spaces between surface and stamp by capillary action. After 30\;min at room temperature, the stamp was removed, and the substrate was rinsed with PBS. Finally, a (50\;$\mu$g/ml) solution of the desired protein (e.g. fibronectin, fibrinogen) dissolved in phosphate-buffered saline (PBS) was added for 30-60\;min (if not noted otherwise) and the substrate was rinsed three times with PBS. 

\subsubsection*{Gradient patterning:}
To set up a protein density gradient in the final pattern, the $\mu$PIPP protocol was applied to a chemotaxis slide.  Therefore, the surface of a chemotaxis slide (ibidi, sticky-Slide Chemotaxis3D) was first patterned with PLL-PEG according to the standard $\mu$PIPP protocol (steps 1 and 2 in Fig.~\ref{fig:1}a). Afterwards, the ``sticky chamber'' was attached and filled with PBS. The PBS on one side of each chamber was then replaced by 45\;$\mu$l of a 100\;$\mu$g/ml solution of protein in PBS according to the manufacturer's instructions \cite{ibidi:chemotaxis} to create a gradient in the concentration of the protein solution. The patterned surface was incubated in this gradient for 40\;min. Finally, the surface was rinsed by flooding the chamber three times with PBS.

\subsubsection*{Multicomponent patterning:}
In order to obtain a pattern consisting of three different components, the basic $\mu$PIPP protocol was extended as shown in Fig.~\ref{fig:1}b. Note that this method works for all stamp geometries that provide enclosed cavities. The PDMS stamp was initially activated for 5\;min in an UV-ozone cleaner (novascan) and incubated for about 1\;h with a 50\;$\mu$g/ml solution of the first protein. Incubated stamps were rinsed once with ultrapure water and dried for about 6\;min. A COC substrate (ibidi) was then activated for 3\;min in the UV-ozone cleaner before the stamp was set in place.\footnote[3]{This UV activation step was found to be critical, since the surface has to be sufficiently hydrophilic to allow the printed protein to be properly transferred from the stamp yet hydrophobic enough to ensure that the protein is adsorbed from the incubation solution.} The subsequent procedure follows the standard $\mu$PIPP protocol. Note that if a third protein instead of PLL-PEG is used, no plasma treatment is necessary.

\subsection*{Cell culture}

\subsubsection*{MDCK:}
The Madin-Darby Canine Kidney (MDCK) cell line was cultured in Minimum Essential Medium (c-c-pro) containing 2 mM L-glutamine and 10\;\% fetal calf serum (FCS) at 37$^{\circ}$C in a 5\;\% $\mathrm{CO_2}$ atmosphere. Prior to experiments, cells were grown to 70-80\;\% confluence, trypsinized and centrifuged at 1000\;rcf for 3\;min. Cell pellets were resuspended in Leibovitz's L15 medium with GlutaMAX (Gibco) and 10\;\% FCS.

\subsubsection*{HUVEC:}
Human umbilical vein endothelial cells (HUVEC) were cultivated in endothelial cell growth medium 2 (ECGM) (Promocell) supplemented with 1\;\% penicillin/streptavidin/amphotericin B (PAN-Biotech) and 10\;\% FCS (PAA). Cells were incubated at 37$^{\circ}$C in a 5\;\% $\mathrm{CO_2}$ atmosphere. Prior to experiments, cells were overlaid with 1x trypsin/ethylenediaminetetraacetic acid (PAN-Biotech) to detach the cell layer. Subsequently, trypsin was inactivated by adding Dulbecco's Modified Eagle's Medium supplemented with 10\;\% FCS. After inactivation, cells were centrifuged and the cell pellet was diluted in ECGM to desired concentration. For all experiments, cells were used in their 3rd passage.

\subsection*{Fluorescence staining}
After a 24\;h incubation on micropatterned plates, cells were washed twice with PBS, and fixed with 4\;\% paraformaldehyde for 10\;min. After a second washing step with PBS, cells were permeabilized for 10\;min using 0.2\;\% Triton in PBS. Before staining, samples were exposed to 1\;\% bovine serum albumin (BSA) in PBS for 30\;min to saturate non-specific protein-binding sites, and then stained with a 1:400 dilution of rhodamine-phalloidin (Invitrogen/Thermo Scientific) and 0.5\;$\mu$g/ml Hoechst 33342 (Sigma) diluted in 1\;\% BSA in PBS. After 30\;min, fixed cells were washed three times for 5\;min each with PBS containing 0.2\;\% BSA and sealed with FluorSave Reagent (Merck Millipore) and a coverslip.

\subsection*{Microscopy}
Phase-contrast and fluorescence images were taken on a Nikon TI Eclipse inverted microscope. Confocal microscopy was performed using a Leica SP8 microscope.

\section*{Results and discussion}

\begin{figure}[ht!]
\centering
  \includegraphics[width=1\linewidth]{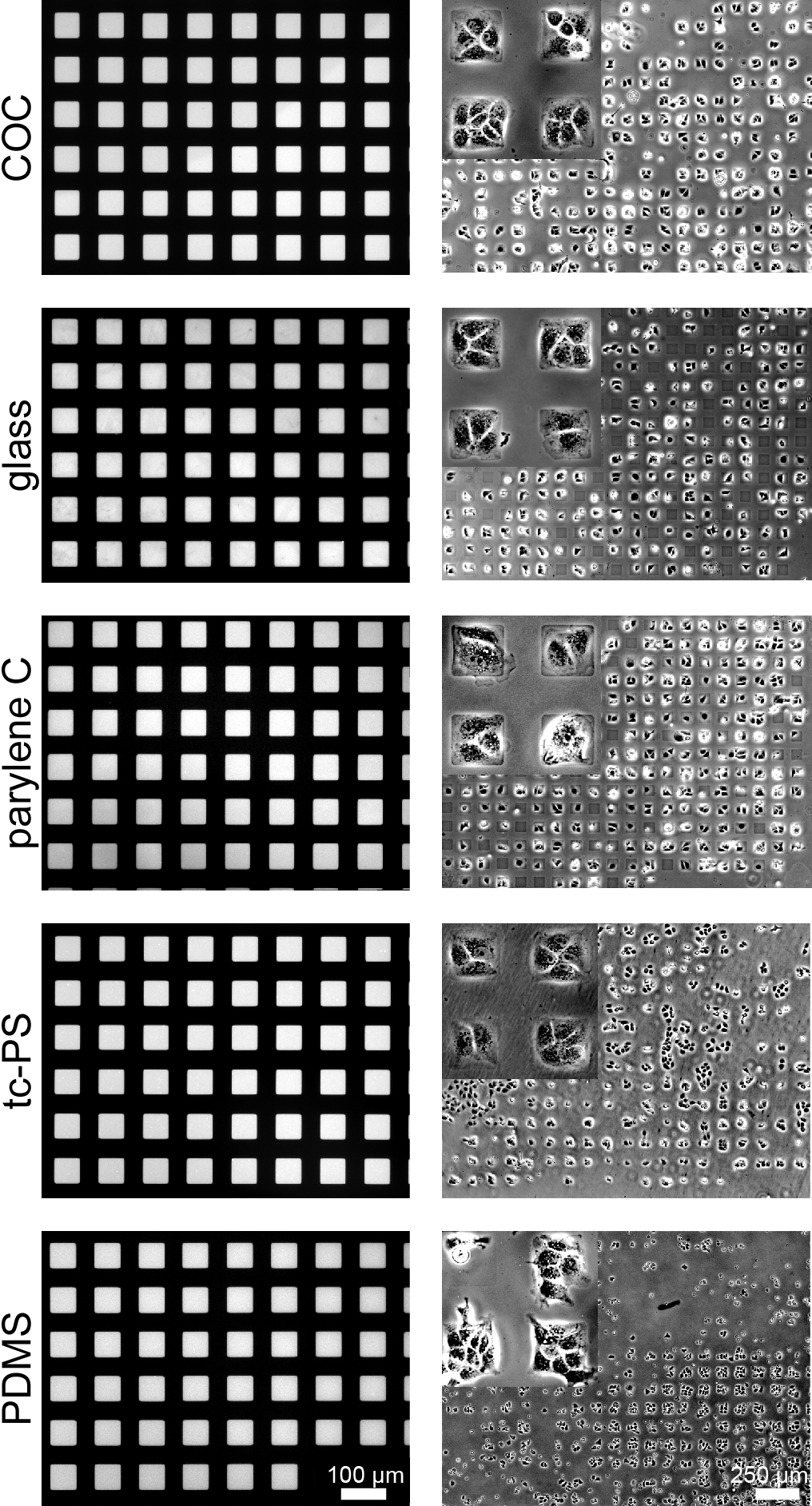}
  \caption{$\mu$PIPP on different substrates. (Left column) Fluorescence images of fibrinogen Alexa-488 patterns on different substrates: COC, glass, parylene C, tc-PS and PDMS. Black areas are passivated with PLL-PEG. (Right column) MDCK cells on fibronectin patterns 24\;h after cell seeding (insets 5x magnified).}
  \label{fig:2}
\end{figure}

In the first set of experiments we created grids of fibronectin-coated squares using microscale plasma-initiated protein patterning ($\mu$PIPP) as described in the Methods section. Note that in this set of experiments we used squares of 60~$\mathrm{\mu m}$ width but in principle pattern resolution is limited by the accuracy of the stamp master ($\sim$2~$\mathrm{\mu m}$ in our experiments). To test whether $\mu$PIPP is compatible with commonly used cell-culture substrates, we applied the procedure to standard tc-PS, glass, COC, parylene C, and PDMS. As depicted in Fig.~\ref{fig:2}, the method produces homogeneous protein patterns on all these surfaces. The method could be applied to these substrates without the need of any surface pretreatment or adjustment of the protocol. Cell adhesion and confinement to the patterned surfaces was achieved on tc-PS, as well as on glass and COC. Pattern quality was also high on parylene C, which is often used as a biocompatible coating for electronic devices. Although patterns were successfully produced on PDMS, confinement of cells within these patterns was not stable over time. (To vary the stiffness of the PDMS surface, monomer to crosslinker ratios of 1:5, 1:10, and 1:20, respectively, were used.)

Notably, patterns produced by treatment with plasma can be visualized by phase-contrast as well as differential interference contrast microscopy (see Fig.S2 of the Supplementary Material), owing to the fact that exposure to plasma slightly alters the altitude and degree of roughness of the surface (in the order of a few hundred nm), and hence changes its optical properties \cite{Beaulieu2009,Alam2014}. This feature simplifies working with the final pattern, as patterned regions can be located and identified without any need for fluorescence microscopy.
In order to test whether the patterns produced are suitable for cell confinement over long timescales, we carried out time-lapse measurements of cells on patterned surfaces over extended time periods. As shown in Fig.S3 of the Supplementary Material (and seen in earlier studies \cite{Roettgermann2014,Ferizi2015,Segerer2015}), the patterns are stable and capable of confining cells over periods of up to several days.

\begin{figure*}[ht]
\centering
  \includegraphics[width=1\linewidth]{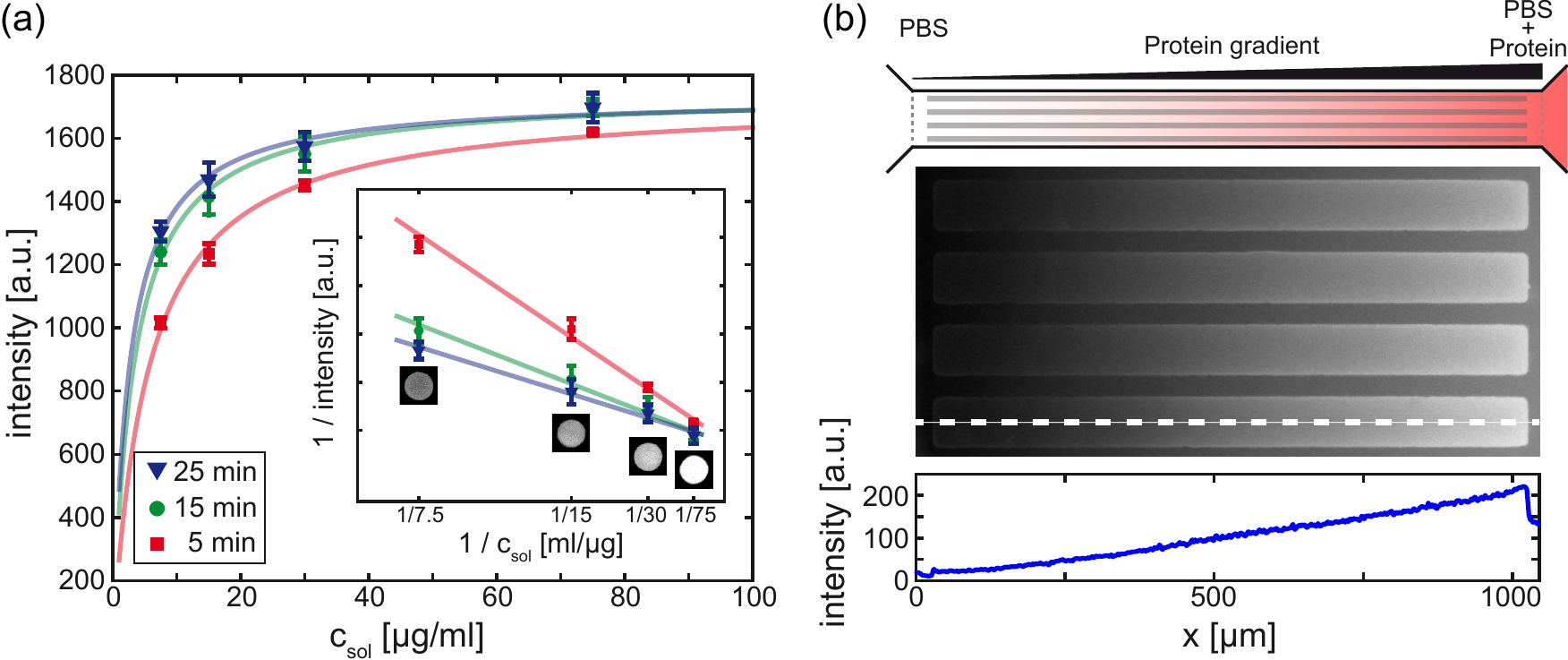}
  \caption{Controlling the surface concentration of the protein. (a) Protein density within patterns can be adjusted by varying both the concentration in the incubation solution $c_{sol}$ and the incubation time $\tau$. We analyzed the fluorescence intensity of Alexa-488-labeled fibrinogen in micropatterns after incubation with $c_{sol}=\;$7.5, 15, 30 or 75\;$\mathrm{\mu g/ml}$ for 5, 15, and 25\;min. The data is well fitted by a Langmuir isotherm (Eq. \ref{eq:langmuir}) with equilibrium constants $\alpha_{\tau}$ of $\alpha_{5}=\;$0.18\;$\pm$0.06, $\alpha_{15}=\;$0.31$\;\pm$0.16, and $\alpha_{25}=\;$0.39$\;\pm$0.18 for the different incubation times $\tau$ respectively ($\pm$ errors indicate confidence bounds of 95\;\% within the fits). The inset shows the corresponding linear scaling in inverse presentation. Error bars indicate the standard deviation. (b) A gradient in the surface-bound protein density can be generated within the patterns by incubation in a protein concentration gradient formed in a chemotaxis chamber. Top down: 1. Formation of gradients on the chemotaxis slide. 2. Fluorescence image of micropatterned stripes obtained by incubation in a gradient of fibrinogen Alexa-488. 3. Measured intensity along the line shown in the middle panel.}
  \label{fig:3}
\end{figure*}

Protein coating by incubation (rather than stamping) has the advantage that the concentration of protein attached to the surface can easily be varied, either by adjusting the concentration of the protein solution $c_{sol}$ or the time of incubation $\tau$. The results of a series of experiments in which both parameters were systematically varied are shown in Fig.~\ref{fig:3}a. Here, for three different incubation times $\tau =\;$5, 15, 25$\;\mathrm{min}$ the concentration of fibrinogen Alexa-488 in the solution was also varied as $c_{sol} =\;$ 7.5, 15, 30, 75$\;\mathrm{\mu g/ml}$. For each combination of $\tau$ and $c_{sol}$, we evaluated the fluorescence intensities of over 1500 patterns from four experiments. The low standard deviation of the measured distribution indicates that very homogeneous and reproducible patterns were generated. The mean values show that an increase in the incubation time $\tau$ or protein concentration in solution $c_{sol}$ leads to an increase in fluorescence intensity, and thus, assuming a linear dependence of protein density to fluorescence intensity, in the surface density of protein.

The adsorption behavior as a function of $c_{sol}$ is well fitted by the Langmuir expression for the adsorption isotherm \cite{Langmuir1918}:
\begin{equation}\label{eq:langmuir}
  c_{surf}=c_{max}\times\frac{\alpha \times c_{sol}}{1+\alpha \times c_{sol}}
\end{equation}
Here, $c_{surf}$ denotes the surface concentration of the protein, $c_{\mathrm{max}}$ the saturated surface concentration, and $\alpha$ the equilibrium constant in the case of Langmuir adsorption. Note, however, that in our experiments adsorption to, and desorption of protein from the surface depends on the time of incubation and hence is clearly not in equilibrium (at least not for $\tau =\;$ 5 and $15\;\mathrm{min}$, as $c_{surf}$ increases further for longer incubation times $\tau$). The data suggests, though, that an equilibrium regime may be asymptotically reached for longer incubation times $\tau$. Still, in the context of this paper, we use the Langmuir expression solely as an estimate for the adsorption behavior for different incubation concentrations. 
The dependence of $c_{surf}$ on $c_{sol}$ can be exploited to generate gradients in the density of the surface-bound protein within the micropatterns. To this end, we used a chemotaxis chamber to create a gradient in the concentration of the protein solution $c_{sol}$ in the protein incubation step of $\mathrm{\mu PIPP}$. Using this method, we succeeded in creating gradients in the density of the surface-bound proteins within the micropatterns, as can be seen in Fig.~\ref{fig:3}b. Such gradients in the concentration of proteins like fibronectin or vascular endothelial growth factor are known to have a guiding effect on cell migration and could therefore be exploited to orchestrate cell motion within the micropatterns \cite{Smith2004,Liu2007}. After the pattern and a surface gradient have been prepared, the chemotaxis chamber can be used to set up an additional gradient of protein in the solution bathing the attached cells. In such a setup, the guidance cues of a gradient in the surface-bound protein density, a soluble protein gradient, and the cues provided by the micropattern are combined. Since \textit{in vivo} cells are often confronted with such multi-cue situations, their implementation \textit{in vitro} is a useful tool for cell sciences \cite{Rodriguez2013}.

\begin{figure*}[ht!]
\centering
  \includegraphics[width=1\linewidth]{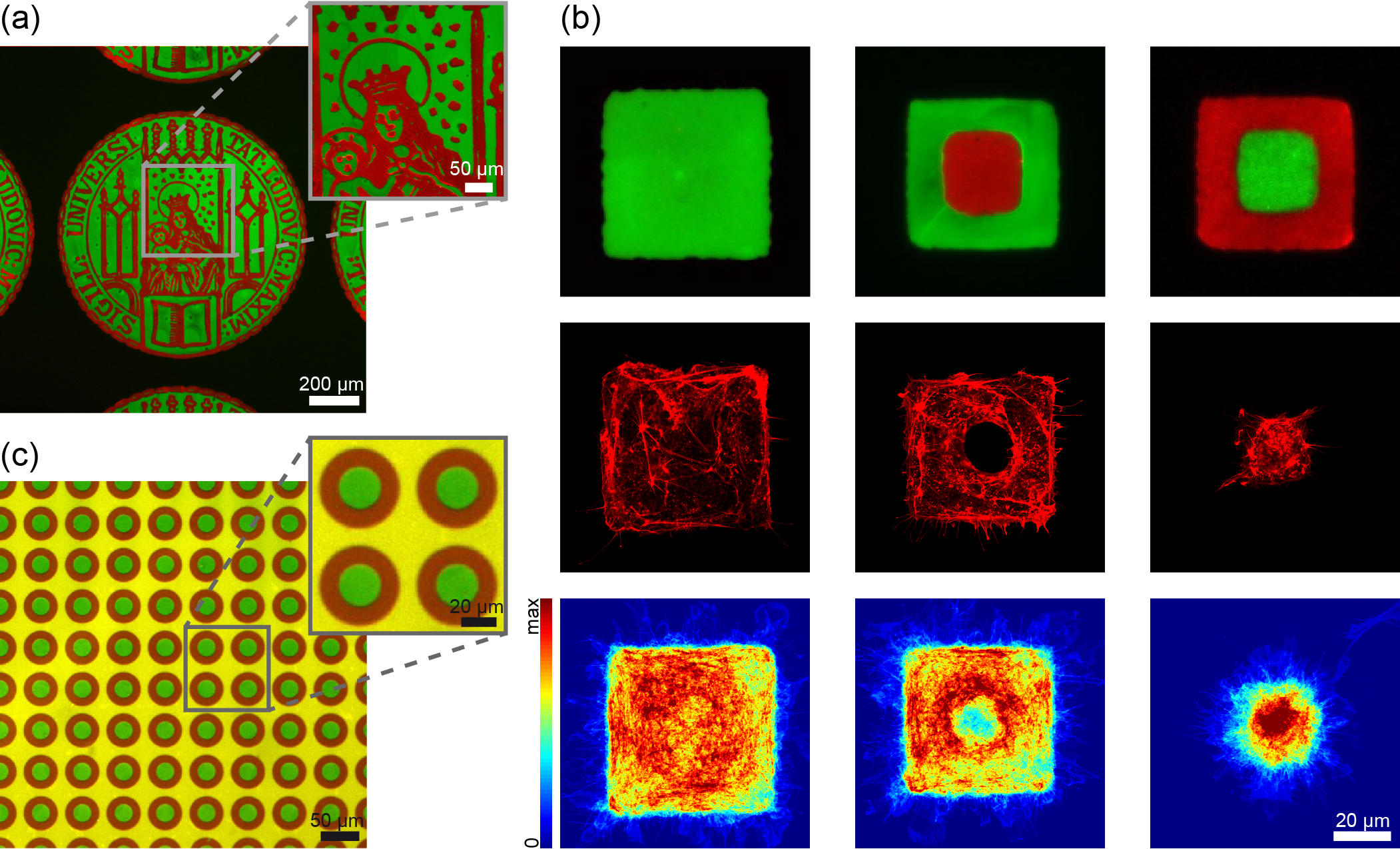}
  \caption{Multicomponent patterning. By combining $\mu$PIPP with $\mu$CP, patterns consisting of three different functionalizations can be formed. (a) A complex pattern consisting of PLL-PEG (black) and fibrinogen labeled with Alexa-488 (green) and Alexa-647 (red), respectively. (b) Fluorescence image of patterns composed of fluorescently labeled fibronectin (green) and laminin (red) (top row). A representative confocal fluorescence image of the actin cytoskeleton of HUVECs arranged in such patterns (middle row), and a heat map of the actin cytoskeleton distribution of cells on over 20 evaluated patterns (bottom row). (Note that all three patterns were created with the multicomponent pattern protocol; hence the inner portions of the framed squares are the only regions available for passive protein attachment by incubation.) (c) Framed circle pattern consisting of fibrinogen labeled with Alexa-488 (green), Alexa-647 (red) and Alexa-594 (yellow), respectively.}
  \label{fig:4}
\end{figure*}

In addition to single protein patterns, the technique is also capable of forming patterns consisting of three components. Such multicomponent patterning can be achieved by combining $\mathrm{\mu PIPP}$ with $\mathrm{\mu CP}$. A simple and versatile implementation of this combination is available for all geometries that are designed in such a way that the PDMS stamp used provides enclosed cavities (Fig.~\ref{fig:1}b). Using such geometries, the embossed parts of the stamp are directly used for $\mathrm{\mu CP}$ whereas the enclosed cavities which shield the surface from plasma without direct stamp contact allow for protein coating according to the standard $\mu$PIPP protocol. In this way, complex multicomponent patterns, such as those depicted in Fig.~\ref{fig:4}a, can be created. Passivation with PLL-PEG is not essential in this procedure, and patterns consisting of three different types of proteins are possible as well (Fig.~\ref{fig:4}c). In contrast to iterative methods of creating patterns of multiple components, the advantage of creating all functionalizations with the aid of the same stamp (and working iteration) is that the individual components can be placed directly adjacent to each other and their relative positioning can hence be accurately controlled. Precise interfaces and pattern geometries consisting of multiple coatings can therefore be guaranteed without the potentially problematic step of bringing one pattern generation in the relatively right position to the last one.
Note that $\mu$PIPP can also be combined with $\mathrm{\mu CP}$ in a successive manner (see Section~S4 of the Supplementary Material). This alternative way of combining both techniques compliments the protocol described above and is able to produce multi component patterns such as ``dashed'' stripe patterns similar to the ones created via stamp-off protocols \cite{Desai2011,Rodriguez2014} (see Fig.~S4 of the Supplementary Material).

By seeding cells on such multicomponent patterns, cellular responses to different surface coatings can be directly compared. As a proof of principle, we studied cell adhesion on ``framed'' patterns consisting of a square area coated with one kind of protein surrounded by a rim area coated with a second type of protein. We chose a standard passivation with PLL-PEG and the two extracellular matrix proteins fibronectin and laminin-1, which are both extracellular matrix proteins known to play a role in cell adhesion \cite{Carlsson1981,Clyman1990}. We found the adhesion of HUVECs to be strongly affected by the different protein coatings within the patterns. The cells were well confined within framed squares. Within the patterns however, they avoided the parts coated with laminin-1, while adhering to the fibronectin-coated areas (Fig.~\ref{fig:4}b). Such decreased or increased adhesion on laminin compared to other proteins has been reported before \cite{Eichinger2012,Carlsson1981}. In our setup, the preference of fibronectin over laminin did neither depend on the pattern geometry nor on which of the proteins was printed and which was applied by incubation, as it becomes evident from comparison of column 2 and 3 in Fig.~\ref{fig:4}b. This suggests that, in this method, the effects of the proteins on cell adhesion do not depend on the way they are affixed to the surface.

\section*{Conclusion}
In this work, we have described $\mu$PIPP, a novel and simple technique for the fabrication of micropatterned protein-coated surfaces for cell studies. As shown, $\mu$PIPP is compatible with various substrates and proteins typically used in cell research. The concentrations of proteins adsorbed to the surface can be readily controlled and gradients in the density of surface-bound proteins can be formed. Both parameters are known to influence on cell spreading and migration \cite{Maheshwari1999,Maheshwari2000,Rajagopalan2004,Smith2004,Liu2007}. Since an additional gradient can be set up in the liquid medium bathing the cells with the aid of the chemotaxis chamber, multi-cue situations can be produced, in which the synergy or competition between surface and solution gradients can be studied within the defined geometries provided by a micropattern. Furthermore, in combination with $\mu$CP, patterns consisting of three different components can be generated. The fact that the deposition areas of all three components emerge from the design of a single stamp brings the advantage of high accuracy in the relative positioning of all components while still maintaining a relatively simple protocol. This set of patterning techniques thus permits complex microenvironments to be created and allows for direct comparisons of the impact of different surface functionalizations on cell adhesion and migration. Due to the simplicity and versatility of the protocol, it should find wide application as a micropatterning tool in cell science labs.

\section*{Acknowledgements}
Financial support from the Deutsche Forschungsgemeinschaft (DFG) via Projects B01 and B08 in Sonderforschungsbereich (SFB) 1032, and from the European Union's Seventh Framework Programme (FP7) for Research (Project NanoMILE) is gratefully acknowledged.






\bibliography{lit} 
\bibliographystyle{rsc} 

\onecolumn
\renewcommand{\thefigure}{S\arabic{section}}
\renewcommand{\thesection}{S\arabic{section}}
\renewcommand{\theequation}{S\arabic{equation}}
\renewcommand{\thetable}{S\arabic{table}}
\thispagestyle{plain}





\vspace{3cm}
\sffamily

\noindent\LARGE{\textbf{``A versatile method to generate multiple types of micropatterns''\\ Supplementary Material}} \\

\noindent\large{F. J. Segerer,\textit{$^{a}$} P. J. F. R\"ottgermann,\textit{$^{a}$} S. Schuster,\textit{$^{b}$} A. Piera Alberola,\textit{$^{a}$} S. Zahler,\textit{$^{b}$} and J. O. R\"adler\textit{$^{\ast a}$}}
\\

\vspace{0.6cm}


\renewcommand*\rmdefault{bch}\normalfont\upshape
\rmfamily
\section*{}
\vspace{-1cm}


\footnotetext{\textit{$^{a}$~Faculty of Physics and Center for NanoScience, Ludwig-Maximilians-Universit\"at M\"unchen, Geschwister-Scholl-Platz 1, D-80539 Munich, Germany. Fax: +49 89 2180 3182; Tel: +49 89 2180 2438; E-mail: raedler@lmu.de}}
\footnotetext{\textit{$^{b}$~Department of Pharmacy - Center for Drug Research, Pharmaceutical Biology, Ludwig-Maximilians-Universit\"at M\"unchen, Butenandtstr. 5-13, D-81377 Munich, Germany }}




\section{Preparation of stamp masters}

As for conventional $\mu$CP, a master of the desired pattern was prepared on silicone using photolithography. An adhesion promoter (TI-Prime, MicroChemicals GmbH) was applied to a silicon wafer (Si-Mat) by spin-coating, first at 500\;rpm for 5\;s and then accelerating to 5000\;rpm for 30\;s. The wafer was baked for 2\;min at 120$^{\circ}$C on a hot plate. Then, a 15\;$\mu$m thick layer of negative photoresist (SU-8 100, micro resist technology GmbH) was applied by spin-coating first at 500\;rpm for 5\;s and then accelerating to 2000\;rpm for 35\;s. Next, the waver was soft baked at 65$^{\circ}$C for 2\;min and then at 95$^{\circ}$C for 5\;min. In the following step, the wafer was exposed to UV light (wavelength peaks: 365\;nm, 405\;nm,\;436 nm) which was passed through a photo mask (e.g. Zitzmann GmbH) of the desired pattern (alternatively, a laser lithography device can be used to write the desired pattern directly in the photoresist). A 5\;min post-exposure baking step was performed at 95$^{\circ}$C in order to selectively crosslink the UV-exposed portions of the resist. Afterwards, the wafer was placed in a developer bath (mr-Dev 600, micro resist technology GmbH) for approximately 2\;min. To inhibit crack formation in the photoresist layer, the wafer was baked again for 5\;min at 95$^{\circ}$C. Finally, the surface was silanized with perfluorotrichlorosilane (Sigma Aldrich) by silane evaporation.

\section{Patterns viewed by phase-contrast and differential interference contrast microscopy}

\begin{figure}[!ht]
\includegraphics[width=1\linewidth]{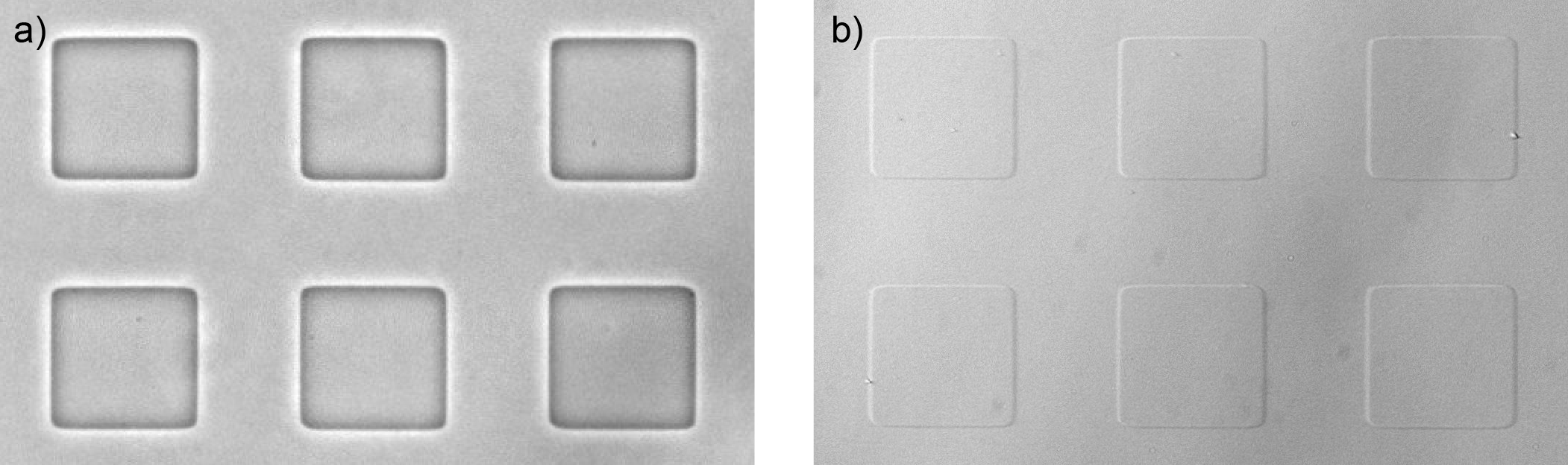}%
\caption{\label{Fig.S.1} 
Microstructures visualized by phase-contrast and differential interference contrast microscopy. A pattern of squares (width 60\;$\mu$m) created on COC by $\mu$PIPP is imaged here by (a) phase-contrast and (b) differential interference contrast microscopy.
}
\end{figure}
\botfigrule

\newpage

\section{Long-term confinement of cells within the patterns}

\begin{figure}[!ht]
\includegraphics[width=1\linewidth]{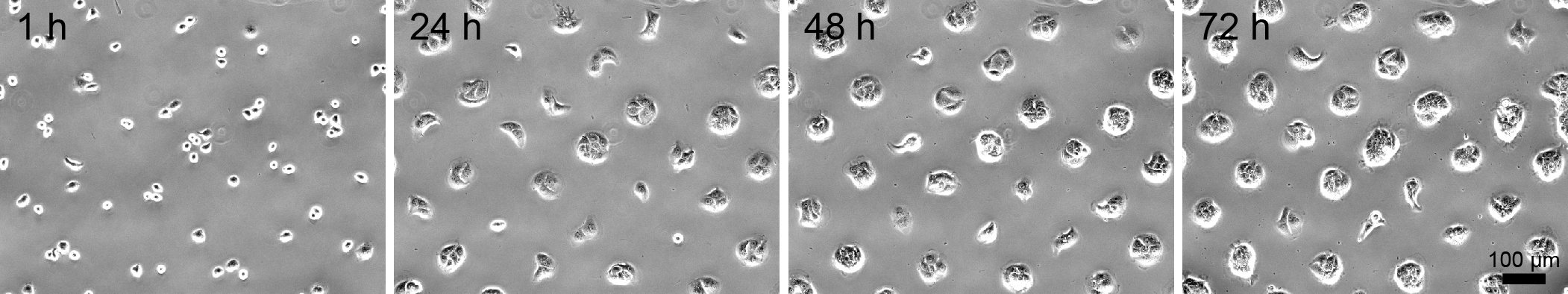}%
\caption{\label{Fig.S.2} 
Patterns are stable and cells are confined to pattern elements for periods of several days. Cell confined in circle shaped micropatterns (radius 46\;$\mu$m) at different time points after cell seeding.
}
\end{figure}
\botfigrule

\section{Multicomponent patterning by successive combination of $\mu$PIPP and $\mu$CP}

\begin{figure}[!ht]
\includegraphics[width=1\linewidth]{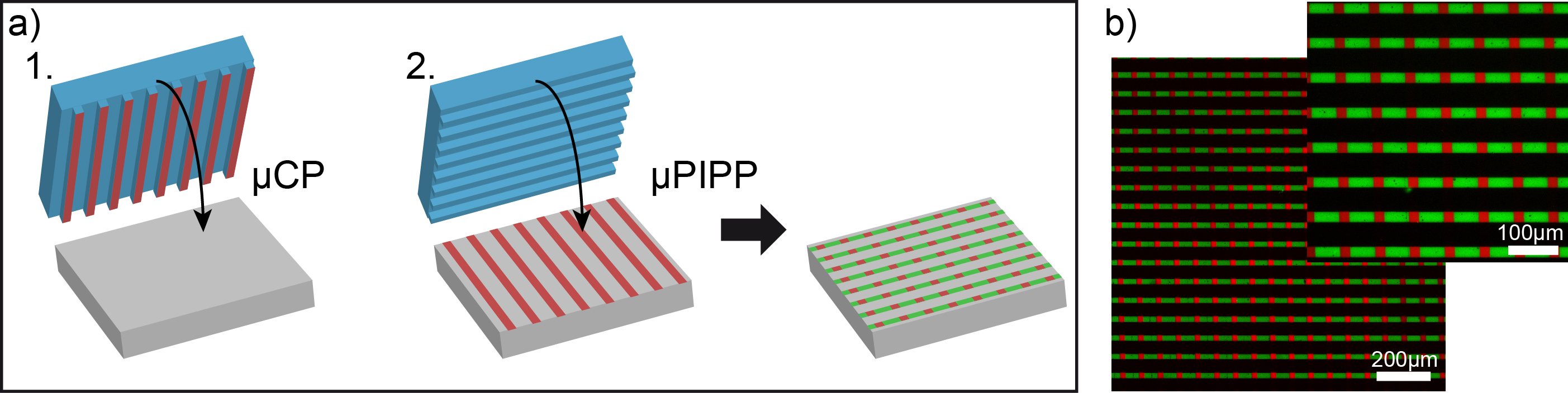}%
\caption{\label{Fig.S.3} 
Alternative combination method of $\mu$PIPP and $\mu$CP. a) Patterning procedure: 1. A pattern of the first protein is created via $\mu$CP. 2. $\mu$PIPP is applied to the already patterned surface such that the pattern created via $\mu$CP is partially removed and replaced by areas coated with a second protein or PLL-PEG respectively. b) Dashed line pattern created via the successive combination of $\mu$CP and $\mu$PIPP. Pattern is composed of PLL-PEG (black), fibrinogen labeled with Alexa-488 (green) and Alexa-647 (red), respectively.}
\end{figure}
\botfigrule

\noindent Additionally to the method described in the main text, $\mu$PIPP can also be combined to $\mu$CP in a successive way to create multicomponent patterns. Here, $\mu$CP and $\mu$PIPP are applied one after another such that the pattern created via $\mu$CP is partially maintained and partially removed by the plasma used for $\mu$PIPP (Fig. \ref{Fig.S.3}a). Analogously to the already described protocol to create multicomponent patterns, a PDMS stamp (in this case a stripe pattern) was initially activated for 5\;min in an UV-ozone cleaner (novascan) and incubated for about 1\;h with a 50\;$\mu$g/ml solution of the first protein. The incubated stamp was rinsed once with ultrapure water and dried for about 6\;min. A COC substrate (ibidi) was then activated for 3\;min in the UV-ozone cleaner before the stamp was set in place. For better transfer of the protein to the surface, a droplet of PBS was applied to the margins of the stamp and drawn underneath by capillary action. The sample was incubated like this for 1\;h. Afterwards, the stamp was removed and the surface dried in an airflow. Now, another stamp (in this case of the same stripe pattern as used above) was placed in such a way on the pattern that the stripes were orthogonal to the stripes created via $\mu$CP. The subsequent procedure follows the standard $\mu$PIPP protocol whereby only the parts of the pattern created via $\mu$CP are preserved that are covered by the second stamp. This successive combination of both techniques can for example be used to create stripe patterns composed of one kind of protein that are interrupted by fields of another protein type (Fig.~\ref{Fig.S.3}b).

\end{document}